\documentstyle[preprint,aps]{revtex}
\tighten
\draft
\begin{document}
\title{Tunneling in One-Dimensional non-Luttinger Electron Liquid}

\author{K. A. Matveev,\cite{*} Dongxiao Yue, and L. I. Glazman}

\address{Theoretical Physics Institute,
School of Physics and Astronomy, University of Minnesota,\\
116 Church Str. SE, Minneapolis, MN 55455}
\maketitle

\begin{abstract}
The conductance of a weakly interacting electron gas in the presense of a
single scatterer is found at arbitrary strength of the scattering potential.
At weak interaction, a simple renormalization group approach can be used
instead of the standard bosonization procedure. Our technique allows to
take into account the {\em backscattering\/} of electrons that leads to a
non-Luttinger-liquid behavior of the low-temperature conductance. In the
presence of magnetic field, the backscattering may give rise to a peak
in differential conductance at bias equal to the Zeeman energy.
\end{abstract}

\narrowtext

Recent progress in semiconductor technology renewed the interest in
transport properties of one-dimensional (1D) electron systems. It is well
known that in a clean short channel the electron transport is ballistic, and
the conductance is quantized \cite{Beenakker}. The ballistic conductance is
destroyed, however, by scattering on impurities in longer channels
\cite{Timp}. It is reasonable to expect that in a sufficiently clean system
it is possible to form a long 1D channel with a single impurity.
Transport properties of such a system are determined by scattering of
electrons on the impurity. In the simplest case of non-interacting
electrons  the conductance is related to the transmission coefficient by
Landauer  formula \cite{Landauer}.

Electron-electron interaction in a 1D electron gas renormalizes
significantly the scattering caused by an impurity potential \cite{Mattis}.
The low-energy properties of a 1D interacting electron system are usually
described by Luttinger model (see, e.g., \cite{Mahan}). This approach allows
to treat the renormalizations of the impurity potential at any strength of
interaction between electrons \cite{Kane}. The resulting conductance $G$
demonstrates the power-law dependence on temperature at $T \to 0$. However,
this method does not allow to calculate the conductance at higher
temperatures where the power-law asymptotics fails. Besides, for the system
of  spin-$\frac12$ electrons, the Luttinger model neglects backscattering
in  electron-electron collisions. The backscattering processes lead to
additional renormalizations that affect the low-temperature conductance.

Below we present an alternative renormalization group (RG) approach that is
valid only in the case of weakly interacting electrons. The advantage of our
approach is in its  ability to describe the conductance at any temperature.
For a model with spinless electrons it allows us to show explicitly the
crossover from the Fermi-gas to the low-temperature Luttinger liquid
behavior. For the realistic case of spin-$\frac12$ electrons, the
backscattering  processes can be incorporated into the renormalization
procedure. Striking differences from Luttinger-liquid behavior occur in the
case of a short-range interaction between electrons. The temperature
dependence of conductance becomes non-monotonic, and application of a
magnetic field creates a peak in the differential conductance at bias
equal to the energy of the Zeeman splitting.

We start with a non-interacting 1D electron gas in the presense of a single
scatterer. In the simplest case of a small-size scatterer we model its
potential by $\delta$-function. The electron scattering is characterized by
the transmission and reflection amplitudes $t_0$ and $r_0$. In terms of
these amplitudes the scattering wave functions have the form:
\begin{equation}
\psi_{k}(x)=\frac{1}{\sqrt{2}}
            \begin{cases}{e^{ikx} + r_0 e^{-ikx}, &  $x<0$,\cr
                          t_0 e^{ikx},            &  $x>0$,}
            \end{cases}
\label{psileft}
\end{equation}
\begin{equation}
\psi_{-k}(x)=\psi_{k}(-x).
\label{psiright}
\end{equation}
Here $\psi_k$ and $\psi_{-k}$ describe the scattering of electrons incoming
from the left and  right correspondingly, $k>0$. Scattering disturbs the
electron density around the barrier. In the presense of electron-electron
interaction this modulation of density leads to an additional scattering of
electrons. If the interaction potential $V(y-z)$ is weak, the correction to
the wave functions (\ref{psileft}) may be found in the Born approximation,
\begin{eqnarray}
\delta\psi_{k}(x)&=&\int_{-\infty}^{\infty}
                 \! G_{k}(x,y)V_{H}(y)\psi_{k}(y) dy \nonumber\\
                 &&-\int\!\!\!\int_{-\infty}^{\infty}
                 \! G_{k}(x,y) V_{\rm ex}(y,z) \psi_{k}(z)dzdy .
\label{born}
\end{eqnarray}
Here $G_k$ is the Green function for non-interacting electrons in the
presense of the barrier. To solve the scattering problem we need only the
asymptotics of the Green function at $x\to -\infty$. In this limit
\begin{equation}
G_{k}(x,y)=\frac{1}{i\hbar v_F}
            \begin{cases}{e^{ik(y-x)} + r_0 e^{-ik(x+y)}, &  $y<0$,\cr
                          t_0 e^{-ik(x-y)},               &  $y>0$,}
            \end{cases}
\label{green}
\end{equation}
where $v_F$ is the Fermi velocity.
The correction (\ref{born}) takes into account the electron scattering
on both Hartree and exchange potentials:
\begin{eqnarray}
V_{H}(y)&=&\int_{-\infty}^{\infty}V(y-z) \delta n(z) dz,
\label{hartree}\\
V_{\rm ex}(y,z)&=&V(y-z)
                 \int_{-k_F}^{k_F}\frac{dk}{2\pi}\psi_{k}^{*}(y)\psi_{k}(z).
\label{exchange}
\end{eqnarray}
At large distances $|z|\gg k_F^{-1}$ the disturbance of density
$\delta n(z)$ caused by the barrier decays as
\begin{equation}
\delta n(z) \simeq \frac{|r_0|}{2\pi|z|}\sin(2k_F|z| - \arg r_0).
\end{equation}
The oscillations of density produce an oscillating Hartree potential
commonly referred to as the {\em Friedel oscillation}, see Fig.~\ref{fig1}.
The asymptotics of  the Hartree potential hints that in Born approximation
the backscattering of  a plain wave should diverge logarithmically at $k\to
k_F$. Indeed, the  explicit calculation based on
Eqs.~(\ref{born})--(\ref{exchange}) gives the  total first-order correction
$\delta {\cal T}$ to the initial value  ${\cal T}_0=|t_0|^2$
of the transmission coefficient
\begin{equation}
\delta {\cal T} = - 2\alpha {\cal T}_0(1-{\cal T}_0)
\ln\left(\frac{1}{|k-k_F|d}\right), \label{first}
\end{equation}
where $d$ is the characteristic spatial scale of the interaction potential.
The dimensionless parameter $\alpha$ characterizes the strength of
interaction:
\begin{equation}
\alpha = \alpha_2 - \alpha_1,
\label{alpha}
\end{equation}
\begin{equation}
\alpha_1 = \frac{V(2k_F)}{2\pi\hbar v_F},\qquad
\alpha_2 = \frac{V(0)}{2\pi\hbar v_F}.
\label{alpha12}
\end{equation}
Here $V(q)$ is the Fourier transformation of the interaction potential. The
two terms proportional to $V(0)$ and $V(2k_F)$ in Eq.~(\ref{alpha})
originate from the exchange and Hartree terms respectively. For the
long-range interaction, $k_F d\gg 1$, the exchange term gives the leading
contribution.

The lowest-order result (\ref{first}) is applicable as long as
$\alpha\ln(1/|k-k_F|d)\ll 1$. At smaller $|k-k_F|$ we have to take into
account the terms of higher orders in $\alpha$. Each term of the
perturbation theory in $\alpha$ diverges at $k\to k_F$. The strongest
divergence in the $n$-th term has the form \mbox{$\alpha^{n}
\ln^{n}(1/|k-k_F|d)$}. We will find the transmission coefficient in the
leading logarithm approximation that corresponds to summation of these most
divergent terms. This task can be performed using a simple renormalization
group approach, analogous to the ``poor man's scaling'' developed by
Anderson \cite{Anderson} for the Kondo problem.

The renormalization of the bare transmission coefficient ${\cal T}_0$
is caused by the interaction with the Fermi sea electrons. Since the
maximum momentum transfer in  a scattering event is determined by the
spatial scale $d$ of the  interaction, only electrons in the energy strip
of halfwidth $D_0=\hbar  v_F/d$ near the Fermi level contribute to the
correction (\ref{first}).  First of all, we neglect the electron states
with energies outside this  strip. Then we transform our problem to a
similiar one with a smaller  bandwidth, $D<D_0$. The two problems are
equivalent if we simultaneously  renormalize ${\cal T}_0$ in order to take
into account the interaction with the  states excluded by this RG
transformation. The renormalization of ${\cal T}_0$  found in Born
approximation, \begin{equation}
\delta {\cal T} = - 2\alpha {\cal T}_0(1-{\cal T}_0)
        \ln\left(\frac{D_0}{D}\right), \label{deltat}
\end{equation}
is similar to Eq.~(\ref{first}). The logarithmic factor corresonds to the
integration over the strip of states excluded by this RG transformation.

We apply the RG transformation again, reducing the bandwidth $D$ step by
step until it reaches the value $|\epsilon|$, where $\epsilon$ is the energy
of the incoming electron counted from the Fermi level. Each step is
accompanied by the renormalization (\ref{deltat}) of the transmission
coefficient. The sum of these small renormalizations may be found as a
solution of the differential equation
\begin{equation}
\frac{d{\cal T}}{d\ln(D_0/D)} = - 2\alpha {\cal T}(1-{\cal T}).
\label{RG}
\end{equation}
The initial condition at $D=D_0$ corresponds to the bare transmission
coefficient ${\cal T}={\cal T}_0$. The renormalized transmission coefficient
for an electron  with energy $\epsilon$ is given by the solution of the RG
equation  (\ref{RG}) at $D=|\epsilon|$,
\begin{equation}
 {\cal T}(\epsilon) = \frac{{\cal T}_0 (|\epsilon|/D_0)^{2\alpha}}
                    {{\cal R}_0+{\cal T}_0(|\epsilon|/D_0)^{2\alpha}},
\label{transmission}
\end{equation}
where ${\cal R}_0=|r_0|^2$ is the reflection coefficient. Formula
(\ref{transmission}) gives the power-law dependence ${\cal T}(\epsilon)$ at
small  $\epsilon$ that coinsides with the result of the Luttinger-liquid
theory   \cite{Kane}. In contrast with this theory, our result
(\ref{transmission})  describes the behavior of transmission coefficient at
all energies   $|\epsilon|<D_0$. Besides, Eqs.~(\ref{alpha}) and
(\ref{alpha12}) gives the   microscopic definition of the exponent $\alpha$
in terms of the interaction   potential. In the case of smooth potential
Eq.~(\ref{alpha}) coinsides  with  the exponent found in \cite{Matveev}.

Our approach allows also to find the temperature dependence of the linear
conductance of a 1D interacting electron system with a single barrier. At
$k_B  T > D_0$ the conductance is given by the Landauer formula for an
ideal Fermi  gas, $G_0=(e^2/2\pi\hbar){\cal T}_0$. At smaller temperatures
the  transmission
coefficient is renormalized. The renormalization of transmission coefficient
should be stopped at  bandwidth $D\sim k_B T$ because of the smearing of the
Fermi surface. As a  result, the following temperature dependence of the
linear conductance is  found:
\begin{equation}
 G(T) = \frac{e^2}{2\pi\hbar}
        \frac{{\cal T}_0 (k_B T/D_0)^{2\alpha}}
                    {{\cal R}_0+{\cal T}_0(k_B T/D_0)^{2\alpha}}.
\label{conductance}
\end{equation}

The differential conductance at a high voltage $eV > k_B T$ may be
obtained by substitution $T\to(e/k_B)V$.

Above we considered a model of spinless electrons.
Experimentally this model can be realized by applying strong magnetic field
that polarizes electron spins. For a system with spin degeneracy (zero
magnetic field) the above theory should be revised. In particular, one has
to take into account the backscattering processes involving two electrons
with different spins. It is well-known \cite{Solyom} that the backscattering
processes cause the renormalizations of interaction constants at low
energies. These renormalizations are commonly studied in the framework of
the following interaction Hamiltonian:
\widetext
\begin{eqnarray}
 H_{\rm int} = \frac{1}{L}\sum_{k,p,q}\sum_{\sigma,\sigma'}\Bigl[&&
 g_{1}
a_{k\sigma}^{\dagger}b_{p\sigma'}^{\dagger}
a_{p+2k_F+q,\sigma'}b_{k-2k_F-q,\sigma}
+g_{2}
    a_{k\sigma}^{\dagger}b_{p\sigma'}^{\dagger}b_{p+q,\sigma'}a_{k-q,\sigma}
\nonumber\\
&&+\frac12
 g_{4}
 \bigl(a_{k\sigma}^{\dagger}a_{p\sigma'}^{\dagger}
       a_{p+q,\sigma'}a_{k-q,\sigma}
 +b_{k\sigma}^{\dagger}b_{p\sigma'}^{\dagger}
  b_{p+q,\sigma'}b_{k-q,\sigma}\bigr)
\Bigr].
\label{hamiltonian}
\end{eqnarray}
\narrowtext
Here $a_{k\sigma}^{\dagger}$ and $b_{k\sigma}^{\dagger}$ are the operators
creating respectively left- and right-moving electrons with momentum $k$ and
spin $\sigma$; interaction constant $g_1$ describes the backscattering, while
$g_2$ and $g_4$ characterize the density-density interaction between the
electrons moving in the opposite directions and in the same direction
correspondingly. Unrenormalized constants $g_1$, $g_2$ and $g_4$ are
determined by the Fourier components of the interaction potential,
\begin{equation}
g_1=V(2k_F),\qquad g_2=g_4=V(0).
\label{constants}
\end{equation}
At low energies the constants $g_1$ and $g_2$ are changing \cite{Solyom} in
the course of renormalization:
\begin{eqnarray}
g_1(D)&=&\frac{V(2k_F)}{1+\frac{V(2k_F)}{\pi\hbar v_F}\ln\frac{D_0}{D}},
\label{17}\\
g_2(D)&=&V(0) - \frac12 V(2k_F) +
              \frac12 \frac{V(2k_F)}{1+\frac{V(2k_F)}{\pi\hbar v_F}
                                                       \ln\frac{D_0}{D}}.
\label{18}
\label{renormalizations}
\end{eqnarray}
Since $g_4$ remains unchanged, the equality $g_2=g_4$ is no longer valid.
Therefore, we have to replace Eq.~(\ref{RG}) with a new RG equation for the
transmission coefficient ${\cal T}$ written in terms of constants $g_1$,
$g_2$, and $g_4$ instead of $V(0)$ and $V(2k_F)$. The first-order
calculation with the Hamiltonian  (\ref{hamiltonian}) gives
\begin{equation}
\frac{d{\cal T}}{d\ln(D_0/D)} = - \frac{(g_2 - 2g_1)}{\pi\hbar v_F}
                          {\cal T}(1-{\cal T}).
\label{newRG}
\end{equation}
Due to the spin degeneracy, the Hartree contribution proportional to $g_1$
contains extra factor of 2 [cf.\ Eq.~(\ref{RG}) and (\ref{alpha})].
Taking into account the dependence (\ref{renormalizations}) of $g_1$ and
$g_2$ on scale $D$ and integrating Eq.~(\ref{newRG}), we find
\begin{equation}
 {\cal T}(\epsilon)=\frac{{\cal T}_0
         \left[1+2\alpha_1\ln\frac{D_0}{|\epsilon|}\right]^{3/2}
         \left|\frac{\epsilon}{D_0}\right|^{2\alpha_2-\alpha_1}}
         {{\cal R}_0+{\cal T}_0
         \left[1+2\alpha_1\ln\frac{D_0}{|\epsilon|}\right]^{3/2}
         \left|\frac{\epsilon}{D_0}\right|^{2\alpha_2-\alpha_1}},
\label{newt}
\end{equation}
where the parameters $\alpha_1$ and $\alpha_2$ are determined by
Eq.~(\ref{alpha12}). Similarly to Eq.~(\ref{conductance}), we can now find
the conductance using the Landauer formula,
\begin{equation}
 G(T) = \frac{e^2}{\pi\hbar}
        \frac{{\cal T}_0 \left[1+2\alpha_1\ln\frac{T_0}{T}\right]^{3/2}
         \left(\frac{T}{T_0}\right)^{2\alpha_2-\alpha_1}}
        {{\cal R}_0+{\cal T}_0\left[1+2\alpha_1\ln\frac{T_0}{T}\right]^{3/2}
         \left(\frac{T}{T_0}\right)^{2\alpha_2-\alpha_1}}.
\label{newconductance}
\end{equation}
The above expression is the main result of this paper. In the absense of
backscattering, $\alpha_1=0$, the existence of the spin degree of freedom
leads only to a trivial factor of 2, as compared to the conductance
(\ref{conductance}) in the spinless case. At $\alpha_1>0$ the backscattering
gives rise to the logarithmic factors in Eq.~(\ref{newconductance}) that
determines the deviation from the power-law asymptotics of conductance at $T
\to 0$. At sufficiently strong backscattering, $\alpha_1 > \frac12\alpha_2$,
the temperature dependence of conductance changes qualitatively. As the
temperature is lowered, the conductance first grows, reaching the maximum
value at
\begin{equation}
   T\sim T_0 \exp\left(-\frac{2\alpha_1-\alpha_2}
                             {\alpha_1(2\alpha_2-\alpha_1)}\right),
\label{crittemp}
\end{equation}
and then drops to zero at $T\to 0$.

The non-monotonic behavior of $G(T)$ is due to the renormalizations induced
by backscattering of electrons with opposite spins. This part of the
electron-electron interaction is responsible for the deviations from the
Luttinger-liquid behavior at all energies. In the presence of magnetic field
$B$ the backscattering is important only at energy scales exceeding
$2\mu_B B$, and the transmission coefficient ${\cal T}(|\epsilon| \geq
2\mu_B B)$ is adequately described by Eq.~(\ref{newt}). At energies below
the Zeeman splitting $2\mu_B B$ the backscattering is suppressed, and the
Luttinger-liquid behavior restores.  The renormalizations of the
transmission coefficient for each spin direction follow the equation
(\ref{RG}) for spinless electrons. The only difference occurs in the
initial conditions that result  from the renormalizations at higher
energies according to (\ref{newt}),
\begin{equation}
{\cal T}_0^{*} \equiv (1-{\cal R}_0^{*}) =
{\cal T}|_{\epsilon = D_0^{*}}, \quad D_0^{*} = 2\mu_B B.
\label{initial}
\end{equation}
Thus the conductance at low temperatures $k_B T \ll \mu_B B$ is described by
Eq.~(\ref{conductance}) with the substitution (\ref{initial}) and with the
extra factor of 2 accounting for the two possible spin directions.

The electron subsystems with the opposite spins decouple at low energies
because the Friedel oscillations produced by these subsystems have
different periods if magnetic field is applied. The two periods correspond
to the two different Fermi wave vectors, $k_{F \uparrow}$ and
$k_{F\downarrow}$. At low temperatures only the electrons with energies close
to the Fermi level contribute to the linear conductance. The low-energy
electron is scattered effectively only by the same-spin Friedel oscillation,
therefore the renormalizations occur in each spin subsystem independently.
However, it is possible to make the electrons to scatter effectively on
Friedel oscillations produced by the opposite-spin subsystem by applying a
finite bias. This scattering shows up as a peak of the differential
conductance at bias $V=2\mu_B B/e$.

At a finite bias $V$ the current $I(V)$ can be expressed in terms of
transmission coefficients ${\cal T}_{\sigma}(\epsilon,V)$ that in turn
depends on voltage,
\begin{equation}
I(V)=\frac{e}{2\pi\hbar}\sum_{\sigma}\int_{-eV}^{0}
     {\cal T}_{\sigma}(\epsilon,V) d\epsilon.
\label{current}
\end{equation}
Here $\sigma=\pm 1$ characterizes the spin direction of tunneling electron,
$\epsilon$ is its energy counted from the Fermi level at the left lead.
We will calculate the transmission coefficient in the limit of strong
barrier, ${\cal T}_{\sigma}\ll 1$. In the presence of bias voltage and
Zeeman splitting, the RG equation (\ref{newRG}) should be modified as follows:
\widetext
\begin{eqnarray}
\frac{d\ln{\cal T}_{\sigma}}{d\ln(D_0/D)}
    &=&- \frac{g_2-g_1}{2\pi\hbar v_F}
         \bigl[\theta(D-|\epsilon|)+\theta(D-|\epsilon+eV|)\bigr]
\nonumber\\
    && + \frac{g_1}{2\pi\hbar v_F}\bigl[\theta(D-|\epsilon-2\sigma\mu_B B|)+
      \theta(D-|\epsilon+eV-2\sigma\mu_B B|)\bigr].
\label{RGlast}
\end{eqnarray}
\narrowtext
We assumed here that the band of the half-width $D$ is centered at the
energy $\epsilon$ of the tunneling electron. The RG transformation
reducing the bandwidth leads to a logarithmic correction to ${\cal T}$ only
as long as the Fermi momentum belongs to the new band. In our case, there
are four relevant momenta corresponding to Friedel oscillations of
electrons with the two spin directions in the two leads, and these
constraints are represented by the four step-functions in the right-hand
side of (\ref{RGlast}). The terms proportional to $g_2 - g_1$ are due
to the Friedel oscillations from the same-spin electrons which contribute
to both exchange and Hartree potentials. The remaining terms are produced
by interaction with the opposite-spin electrons.

The solution of RG equation (\ref{RGlast}) at $D \to 0$, found neglecting
the renormalizations of the interaction constants \cite{remark}, has the
form:
\begin{eqnarray}
{\cal T}_{\sigma}(\epsilon) &=& {\cal T}_0
\left|\frac{\epsilon}{D_0}\right|^{\alpha_2 - \alpha_1}
\left|\frac{\epsilon+eV-2\sigma\mu_B B}{D_0}\right|^{-\alpha_1}
\nonumber\\
&&\times
\left|\frac{\epsilon+eV}{D_0}\right|^{\alpha_2 - \alpha_1}
\left|\frac{\epsilon-2\sigma\mu_B B}{D_0}\right|^{-\alpha_1}.
\label{Tlast}
\end{eqnarray}
This result combined with Eq.~(\ref{current}) allows to calculate the
$I$-$V$ characteristic. As it is clearly seen from Eq.~(\ref{Tlast}),
two singularities in ${\cal T}_\sigma(\epsilon)$ merge at $eV=2\mu_B B$.
As a result a singularity appears in the $I$-$V$ characteristic. The latter
manifests itself as a peak in differential conductance,
\begin{equation}
\delta G=\frac{e^2}{\pi\hbar}{\cal T}_0^{*}
       \frac{\alpha_1}{\alpha_2-2\alpha_1}\left(1 -
       \left|\frac{eV-2\mu_B B}{2\mu_B B}\right|^{\alpha_2-2\alpha_1}\right),
\label{Glast}
\end{equation}
schematically shown in Fig.~\ref{fig2}.
The last result was obtained in the linear in ${\cal T}$ approximation that
accounts for the processes in which electron crosses the tunnel
barrier only once. This approximation is equivalent to the standard
method of calculation of tunnel current in terms of the transmission
coefficient and tunnel densities of states in the leads. Indeed, the
expression (\ref{Tlast}) may be interpreted as the product of the bare
transmission coefficient ${\cal T}_0$ and the power-law factors
corresponding to the singular energy dependence of the tunnel densities of
states. Apparently this method fails if the differential conductance
becomes of the order of $e^2/\hbar$. This does not happen for a
sufficiently long-range interaction potential, $g_2 -2g_1 >0$. In the case
of a short range potential, $g_2 -2g_1 < 0$, formula (\ref{Glast}) is not
applicable in a very narrow vicinity of the point $V=2\mu_BB/e$.

In conclusion, we solved the problem of scattering on a single impurity
for weakly interacting 1D electrons. This allowed us to find the system
conductance at any temperature for spinless case (\ref{conductance}) and
for the real spin-$\frac12$ electrons (\ref{newconductance}) as well.
The electron-electron backscattering gives rise to deviations from the
power-law temperature dependence of linear conductance at $T\to0$.
In the presence of a magnetic field, the backscattering creates a peak
(\ref{Glast}) in the differential conductance at bias $V=2\mu_B B$.

The authors are grateful to H.~U. Baranger, C.~L. Kane, A.~I. Larkin, and
B.~Z. Spivak for helpful  discussions. This work was supported  by NSF Grant
DMR-9117341.

\begin{figure}
\caption{Total scattering potential $V(x)$. The central peak is the bare
potential of the barrier. The wings represent the Friedel
oscillation induced by the barrier.}
\label{fig1}
\end{figure}

\begin{figure}
\caption{The differential conductance $G(V)=dI/dV$ of a quantum wire with
a strong barrier in the presence of magnetic field $B$. The power-law
behavior at $V \to 0$ is consistent with the Luttinger-liquid theory
\protect\cite{Kane}. The power-law peak (\protect\ref{Glast}) at
$V=2\mu_BB/e$ is due to the scattering of electrons on the Friedel
oscillation produced by the opposite-spin electrons. The curve was
calculated for $\alpha_1=1/16$, $\alpha_2 = 3/8$.}
\label{fig2}
\end{figure}


\begin{references}
\bibitem{remark} This approximation is valid for small $g_1$, when
$g_1\ln(D_0/\mu_B B)\ll 1$. If the latter condition is not satisfied, the
result (\ref{Glast}) should be corrected by the substitution of the
renormalized values (\ref{17}) and (\ref{18}) of $g_{1,2}$ at $D=2\mu_B B$
into the definition $\alpha_{1,2}=g_{1,2}/2\pi\hbar v_F$.
\end{references}
\end{document}